\input{aipcheck.tex}

\newcommand\sps{\space\space\space\space}
  \def\selectedoptions{final}

\documentclass[
   \selectedoptions
  ]
  {aipproc}

\layoutstyle{8x11double}

\SetInternalRegister\hbadness{8000} 

\newcommand\doingARLO[2][]{%
  \ifx\mmref\undefined #1\else #2\fi
}

\begin{document}

\title 
      [Clues to unveil the emitter in LS~5039]
      {Clues to unveil the emitter in LS~5039:\\ 
      powerful jets vs colliding winds}


\author{V. Bosch-Ramon}{
  address={Max Planck Institut f\"ur Kernphysik, Saupfercheckweg 1, Heidelberg 69117, Germany},
  email={vbosch@mpi-hd.mpg.de},
  thanks={This work was commissioned by the AIP}
}

\iftrue
\author{D. Khangulyan}{
  address={Max Planck Institut f\"ur Kernphysik, Saupfercheckweg 1, Heidelberg 69117, Germany},
  email={dmitry.khangulyan@mpi-hd.mpg.de},
}

\author{F. A. Aharonian}{
  address={Dublin Institute for Advanced Studies, Dublin, Ireland},
  email={Felix.Aharonian@mpi-hd.mpg.de},
  homepage={},
  altaddress={Max Planck Institut f\"ur Kernphysik, Saupfercheckweg 1, Heidelberg 69117, Germany}
}
\fi

\copyrightyear  {2001}

\begin{abstract}
LS~5039 is among the most interesting VHE sources in the Galaxy. Two scenarios
have been put forward to explain the observed TeV radiation: jets vs pulsar
winds. The source has been detected during the superior conjunction of
the compact object, when very large gamma-ray opacities are expected. In
addition, electromagnetic cascades, which may make the system more transparent
to gamma-rays, are hardly efficient for realistic magnetic fields in massive
star surroundings. All this  makes unlikely the standard pulsar scenario for LS~5039, 
in which the emitter is the region located between the star and the compact object, where the
opacities are the largest. Otherwise, a jet-like flow can transport energy to
regions where the photon-photon absorption is much lower and the TeV radiation is
not so severely absorbed.
\end{abstract}

\date{\today}

\maketitle

\section{Introduction}

LS~5039 is a high-mass X-ray binary located at distances of 2.5~kpc \cite{Casares et
al. 2005} that presents extended non-thermal radio emission \cite{Paredes et al.
2000}. This source has been detected in the very high-energy (VHE) range all
along the orbit and also during the superior conjunction of the compact object
\cite{Aharonian et al. 2005,Aharonian et al. 2006}, when the photon-photon absorption is expected to be
the strongest \cite{Dubus 2006a}.

The primary object in LS~5039 is an O star, and the compact object is still of unknown nature \cite{Casares et al. 2005}. 
Although thought to be a microquasar after the discovery of the extended radio emission \cite{Paredes et al. 2000}, the non
detection of accretion features in the X-ray spectrum led \cite{Martocchia et al. 2005} to suggest the presence of a
non-accreting pulsar in the system. More recently, the nature of the non-thermal emitter has been discussed in several works
(e.g. \cite{Dubus 2006b,Bosch-Ramon et al. 2007,Ribo et al. 2008}), the discussion focusing basically in two possible energy
providers for the VHE radiation: a supersonic outflow or jet ejected from the surroundings of the compact object (e.g.
\cite{Paredes et al. 2006}); or an ultrarelativistic wind produced by a non-accreting pulsar (e.g. \cite{Dubus 2006b}). 

In this work, we study in detail the photon-photon absorption and its consequences, and point out that, under reasonable
values for the ambient magnetic field, the TeV emitter should be located in the borders of the binary system at least around
the superior conjunction of the compact object. Otherwise, the electron-positron pairs created via photon-photon absorption
in the stellar photon field could radiate overcoming the observed broadband emission levels. These pairs can radiate as well
all along the orbit, and may be dominant over a primary component of electrons (see also \cite{Bosch-Ramon et al. 2008}). We
remark that, in \cite{Bosch-Ramon et al. 2007},  it was suggested that the X-ray emitter in LS~5039 may be far from the star,
given the negligible inferred soft X-ray absorption in the stellar wind. In addition, in \cite{Khangulyan et al. 2008}, from
acceleration efficiency arguments, it was proposed that the VHE emitter should be in the borders of the binary system.

\begin{figure}
\caption{Absorbed luminosity $\times 1/4\pi d_{\odot}^2$, 
depending on the location of the emitter in the binary system. The contour line limits the
region to the left, in which the emitter, if located there, would overcome the observed X-ray fluxes if the synchrotron 
emission is the dominant radiation channel.}
\includegraphics[height=.5\textheight]{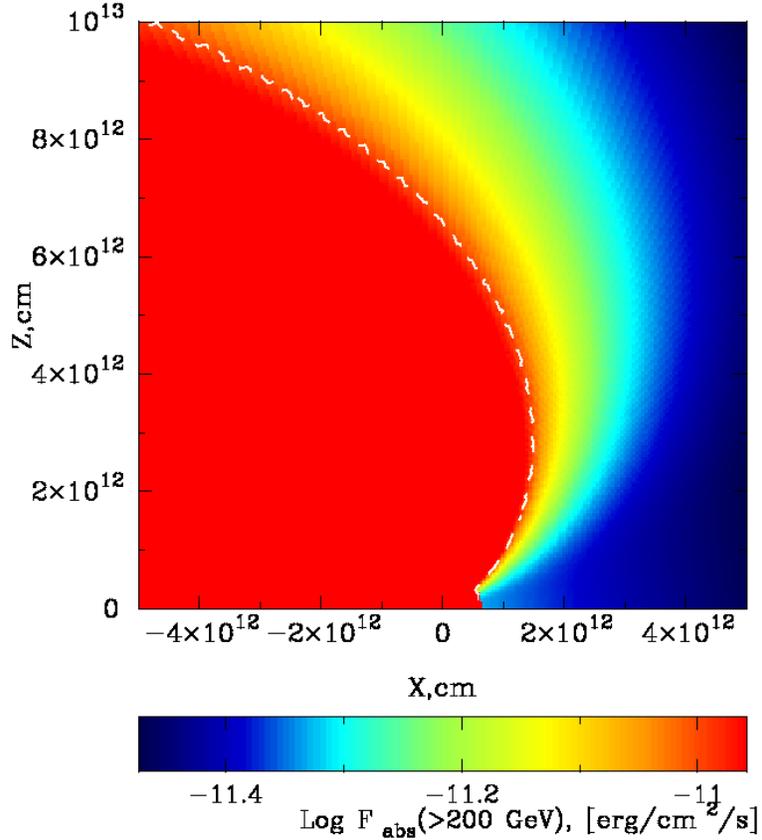}
\label{fig1}
\end{figure}

\begin{figure}
\caption{Computed spectral energy distribution of the synchrotron emission produced by the secondary pairs created in the
surroundings of the VHE emitter during the superior conjunction of the compact
object. The emitter location has been taken close to the compact object, with a
magnetic field of 10~G, and an inclination angle of $60^{\circ}$.}
\includegraphics[height=.25\textheight]{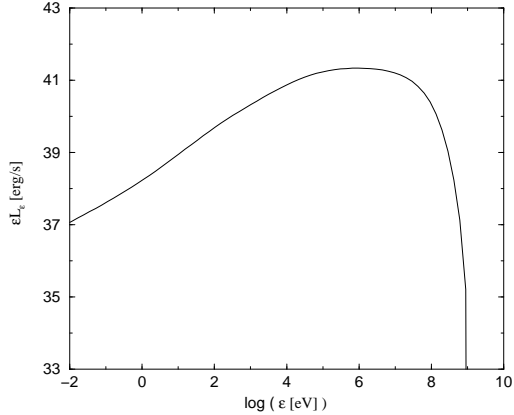}
\label{fig2}

\end{figure}
\begin{figure}
\caption{Top: Computed SED of the synchrotron and IC 
emission produced by primary electrons in LS~5039 around inferior conjunction of the compact
object accounting for photon-photon 
absorption. The HESS data points are also shown. Bottom: Computed SED of the 
synchrotron and IC emission by secondary pairs in 
LS~5039 around inferior conjunction of the compact
object.}
\includegraphics[height=.6\textheight]{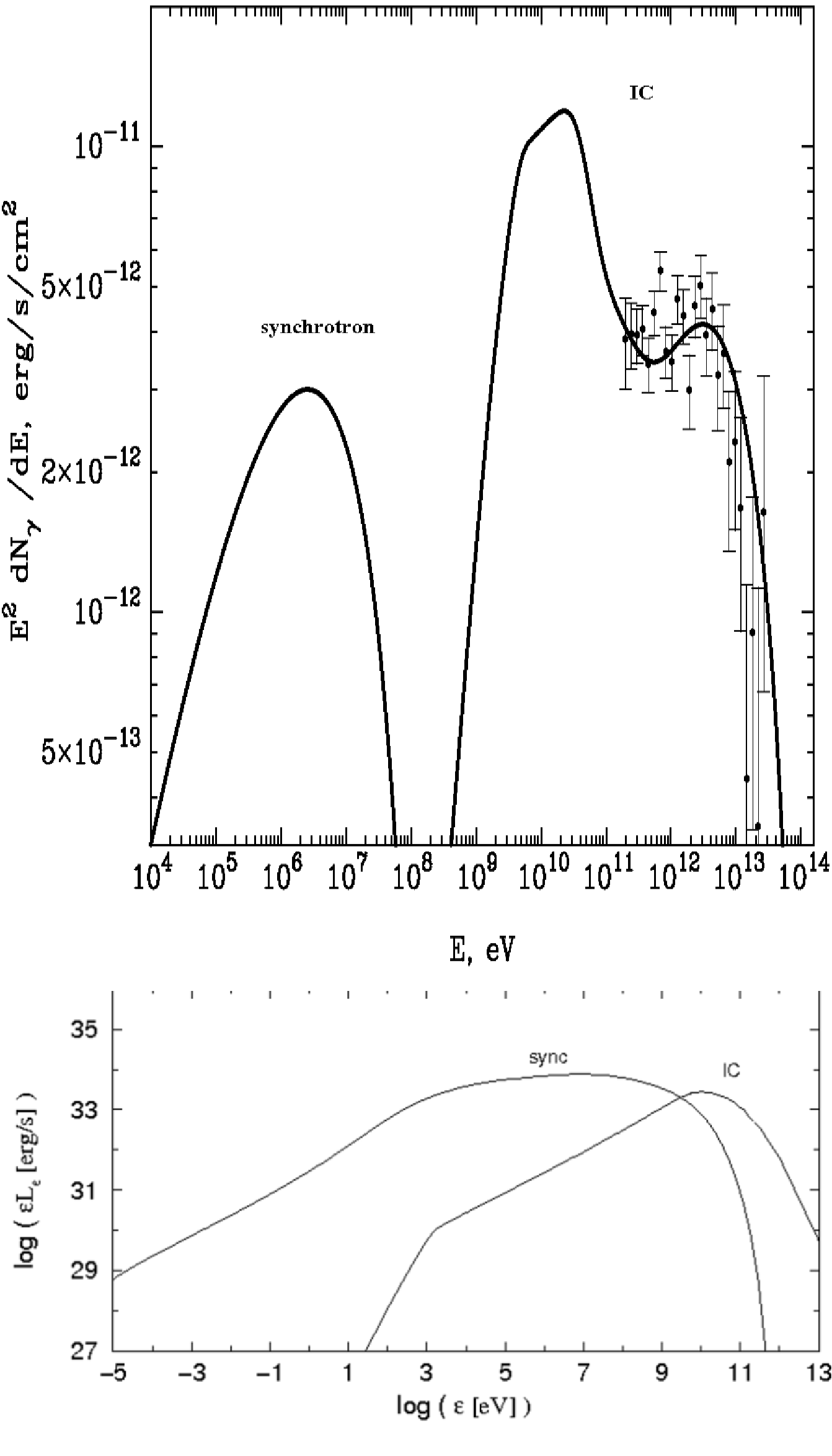}
\label{fig3}
\end{figure}

\section{The TeV emitter in LS~5039}

We have plotted a 2-dimensional map of the absorbed luminosity via photon-photon absorption 
depending on the location of the emitter in LS~5039. This is
presented in Figure~\ref{fig1}. The XZ coordinates correspond to the plane formed by the emitter and the star (0,0)
positions, and the observer line of sight. To compute these maps, we have deabsorbed the observed spectra and fluxes >100~GeV
from LS~5039 around the superior conjunction of the compact object. The regions forbidden by the X-ray observational
constraints are limited by a contour line.  The VHE emitter cannot be located to the left of the dashed line or the X-ray
observed fluxes will be overcome by the secondary pair synchrotron emission, expected to be the dominant secondary pair cooling channel
for reasonable magnetic field values in the stellar surroundings. O-star surface magnetic fields may reach $\sim$~kG
\cite{Donati et al. 2002}, and values of $\ge 10$~G could be realistic at few stellar radii from the stellar surface, more
than enough to suppress electromagnetic cascades \cite{Khangulyan et al. 2008}. In the plot, the compact object is located in
the left half of the XZ plane at $1.4\times 10^{12}$~cm from the star independently of the orbital inclination.

In Fig.~\ref{fig2}, we show the spectral energy distribution (SED) of the synchrotron emission produced by the secondary
pairs created in the surroundings of the VHE emitter during the superior conjunction of the compact object. The emitter
location has been taken close to the compact object, with an ambient magnetic field of 10~G, and an inclination angle of
$60^{\circ}$. This value for the inclination angle would correspond to the case of a neutron star as the compact object
\cite{Casares et al. 2005}, e.g. a pulsar. As seen in the figure, the resulting X-ray luminosity is seven orders of magnitude
larger than the one found by observations \cite{Bosch-Ramon et al. 2007}. As noted in Fig.~\ref{fig1}, only in case the
emitter were located far from the compact object, and far as well from the line joining the compact object and the star, the
synchrotron radiation would not overcome the observed fluxes. This is consistent with the fact that the X-rays may come from
the system borders, as suggested in \cite{Bosch-Ramon et al. 2007}.

When the system is in the inferior conjunction of the compact object, the opacities in the direction to the observer are low,
implying a smaller energy budget to power the TeV emission, but the absorbed energy in all the directions is still relatively
large. In Fig.~\ref{fig3}, we show the SEDs of  the synchrotron and inverse Compton (IC) emission of the primary electrons
(top) and the secondary pairs (bottom) created by photon-photon absorption. The primary (absorbed) IC component reproduces
the spectrum detected during the inferion conjunction of the compact object. The fluxes can be easily converted to
luminosities multiplying by $\approx 10^{45}$~cm$^2$. The X-ray secondary pair luminosities could be even larger than those
produced by the primaries. The produced pairs, for reasonable magnetic fields, would generate efficiently synchrotron
emission with fluxes similar to those observed in X-rays.  The secondary pair IC component would contribute to the GeV
range. 

\section{Conclusions}

In summary, the detection by HESS of VHE radiation from LS~5039 during the
superior conjunction of the compact object, plus a realistic value of the
ambient magnetic field, point strongly to an emitter located far away from the
compact object, and far as well from the region between the
pulsar and the star. This cannot discard a more
general pulsar scenario (e.g. \cite{Bogovalov et al. 2008}), but disfavors any pulsar
scenario in which the emission would come from the region between the star
and the pulsar, as in those proposed for LS~5039 to date
(e.g. \cite{Dubus 2006b,Dubus et al. 2008,SierpowskaTorres2007,SierpowskaTorres2008}). 
A jet-like emitter scenario, whatever the formation mechanism,
is favored, given the capability of jets to transport energy and radiate it at
large distances from the compact object and the star.

\begin{theacknowledgments}
V.B-R. gratefully acknowledges support from the Alexander von Humboldt
Foundation. V.B-R. acknowledges support by DGI of MEC under grant
AYA2007-68034-C03-01, as well as partial support by the European Regional
Development Fund (ERDF/FEDER).
\end{theacknowledgments}

\end{document}